\documentclass[compsoc, conference, a4paper, 10pt, times]{IEEEtran}

\usepackage{eso-pic}
\usepackage{tikz}
\usepackage{orcidlink}
\usepackage{cite}
\usepackage{amsmath,amssymb,amsfonts}
\usepackage{algorithm}
\usepackage{algpseudocode}
\usepackage{graphicx}
\usepackage{textcomp}
\usepackage{xcolor}
\usepackage{booktabs}
\usepackage{xspace}
\hypersetup{hidelinks}
\usepackage{tablefootnote}

\newcommand{\tool}[1]{\textsf{\small{#1}}\xspace}
\newcommand{\coreml}{\tool{Core~ML}}
\newcommand{\detr}{\tool{DETR}}
\newcommand{\metal}{\tool{Metal}}
\newcommand{\screncapturekit}{\tool{Screen Capture Kit}}

\newcommand{\macos}{macOS\xspace}
\newcommand{\etal}{\textit{et al.}\xspace}

\AddToShipoutPictureBG*{%
    \AtPageUpperLeft{%
        \setlength{\unitlength}{1in}%
        \begin{picture}(0,0)%
            \put(0.85,-0.75){%
                \begin{tikzpicture}
                    \node[draw, rectangle, thick, text width=\dimexpr\textwidth-0.2in\relax, align=center] {
                        \textbf{Extended version of the manuscript accepted to STAST'24}
                    };
                \end{tikzpicture}
            }%
        \end{picture}%
    }%
}

\begin{document}

\title{Position Paper: Think Globally, React Locally --- Bringing Real-time Reference-based Website Phishing Detection on \macos}

\author{\IEEEauthorblockN{Ivan Petrukha \orcidlink{0009-0009-3427-6869}}
\IEEEauthorblockA{\textit{MacPaw} \\
Kyiv, Ukraine \\
petrukha@macpaw.com}
\and
\IEEEauthorblockN{Nataliia Stulova \orcidlink{0000-0002-6804-2253}}
\IEEEauthorblockA{\textit{MacPaw} \\
Kyiv, Ukraine \\
nata.stulova@macpaw.com}
\and
\IEEEauthorblockN{Sergii Kryvoblotskyi \orcidlink{0009-0007-8006-546X}}
\IEEEauthorblockA{\textit{MacPaw} \\
Kyiv, Ukraine \\
krivoblotsky@macpaw.com}
}

\maketitle

\begin{abstract}%
\paragraph{\textbf{Background.}~%
The recent surge in phishing attacks keeps undermining the effectiveness of the existing anti-phishing approaches.
While state-of-the-art phishing detection solutions can effectively detect phishing across the web, they mainly focus on automated web crawling and updating blacklists%
}
\paragraph{\textbf{Aim.}~%
The time after a phishing website is published on the web and remains unprocessed by detection systems is critical.
We aim to reduce this time gap via real-time phishing detection solution and perform continuous background processing without extra user interaction%
}
\paragraph{\textbf{Method.}~%
We propose a real-time on-device solution that identifies phishing sites immediately when encountered by the user.
Our reference-based approach analyzes the visual content of webpages, identifying phishing attempts through layout analysis, brand impersonation recognition, and credential input areas detection%
}
\paragraph{\textbf{Results.}~%
Our case study shows that it's feasible to perform background processing on-device continuously.
For web browser phishing detection, the process utilizes 16\% of a single CPU core and less than 84MB of RAM on an Apple M1 while maintaining a high efficiency with 95.7\% precision and 87.7\% recall, based on a test dataset of 50K phishing and benign webpages%
}
\paragraph{\textbf{Conclusions.}~ %
Our results demonstrate the potential of on-device, real-time phishing detection systems to enhance cybersecurity defensive technologies.
We maintained the accuracy of state-of-the-art reference-based phishing detection solutions while bringing this functionality directly to the device%
}
~
\end{abstract}


\section{Introduction}
According to the phishing activity trends reports\footnote{\url{https://apwg.org/trendsreports/}} from the Anti-Phishing Working Group (APWG), the number of phishing attacks keeps increasing year over year, with 2023 being the worst so far,
with 4,987,809 unique phishing webpages created.
While cloud infrastructure providers like Cloudflare implement phishing detection on their end\footnote{\url{https://blog.cloudflare.com/2023-phishing-report}},
these measures are not sufficient to prevent phishing attacks at large.
Existing phishing detection methods can be categorized into three categories: \emph{blacklist-based}, \emph{classification-based}, and \emph{reference-based}.
Traditional blacklist-based solutions like \tool{Google Safe Browsing}\footnote{\url{https://safebrowsing.google.com}}, although they are effective, do not keep up with the speed of phishing websites spreading~\cite{2020PhishTime}, whose creation and deployment are typically automated.
Classification-based approaches~\cite{marchal_know_2016,marchal_off--hook_2017,le_urlnet_2018} use machine learning algorithms to analyze URL, HTML, or other features to classify webpages as phishing or legitimate, including on-device approaches~\cite{dalgic_phish-iris_2018,armano_real-time_2016}, but lose efficiency against HTML obfuscation techniques and lack accuracy.
In contrast, reference-based approaches~\cite{abdelnabi_visualphishnet_2020,lin_phishpedia_2021,liu_inferring_2022} use computer vision and can effectively analyze webpage appearance to extract information about the displayed content, allowing to detect phishing attempts and form reasonable verdicts.

While the latest reference-based solutions effectively detect phishing across the web, they focus on automated web crawling and filling the blacklists.
The time after a phishing webpage is published on the web and remains unprocessed by detection systems is critical.
A phishing campaign starts spreading malicious URLs just when they become available, and even within a few minutes, the campaign may affect many users.

In this paper, we propose a real-time on-device unobtrusive anti-phishing solution for \macos and evaluate in on the case of in-browser phishing website detection, improving upon the work of~\cite{liu_inferring_2022}.
Our approach relies on \macos-specific system resources and frameworks, which allows us to significantly reduce computational resource demand.
Processing live screen capturing requires applying a phishing detection algorithm within a matter of seconds, ensuring users' safety.
To achieve this, we rely solely on local machine learning models, eliminating the need for cloud-based tools.
Another important aspect of using only local models is privacy concerns. 
By utilizing local models, user data remains on the device at all times, ensuring enhanced security and peace of mind for the user, which aligns with Apple's approach to privacy protection\footnote{\url{https://www.apple.com/ua/privacy/approach-to-privacy/}}.

\section{Threat Model}
\vspace{-6mm}

\begin{center}
\begin{figure*}[ht!]
  \includegraphics[width=1.0\textwidth]{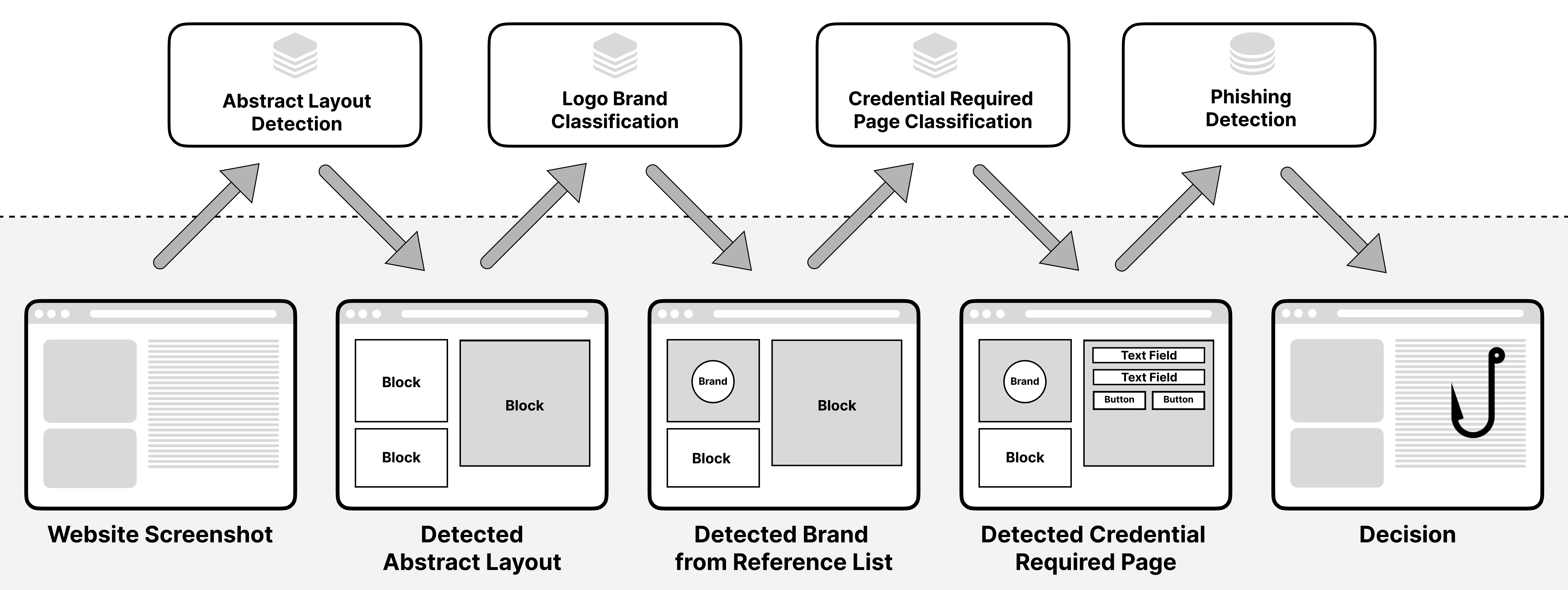}
  \caption{General workflow of our phishing detection approach}
  \label{fig:solution-workflow}
\end{figure*}
\end{center}

We consider the threat model of PhishIntention~\cite{liu_inferring_2022}, in which the adversary creates a phishing website, impersonating an official website of a known company and demanding the personal data of the website visitor.
This form of phishing is executed through social interaction, employing psychological manipulations to trick users into disclosing sensitive security data.
Initially, the attacker conducts a thorough analysis to identify the potential vulnerabilities within the target demographic essential for the successful execution of the attack.
Subsequently, the phisher endeavors to establish trust with the target.
In the final phase, the attacker manipulates the scenario to induce the target to share critical information.

\section{Solution}
\label{sec:solution}

Figure \ref{fig:solution-workflow} presents a high-level overview of our phishing detection approach.
Similarly to PhishIntention, we focus on identifying both the intention to impersonate a brand and the intention to capture user personal information.
The main algorithm consists of three essential steps.
First, we begin with the webpage screenshot analysis (Section~\ref{ssec:abstract_layout_detection}) to identify layout elements and gain rich knowledge about webpage content.
Second, we classify the detected logo brand (Section~\ref{ssec:logo_brand_classification}) that the phishing webpage tries to impersonate. 
If we find a brand from our reference list, we continue to the next step.
Third, we classify the webpage into two categories: if it requires credentials or not (Section~\ref{ssec:crp_classification}).
This step confirms that an impersonated phishing website is trying to steal personal user information.
Matching these conditions allows us to decide whether a webpage is phishing or legitimate.

\subsection{Abstract Layout Detection}
\label{ssec:abstract_layout_detection}

The main purpose of abstract layout detection is to determine whether the currently visible webpage contains a brand logo that may be targeted by hackers and find credential input forms.
To achieve this, we analyze screenshots, identifying different layout elements such as logos, buttons, inputs, labels, and blocks.
We used the part of the dataset from the section~\textit{CRP classifier and AWL detector} of~\cite{phishintention_dataset}.
The dataset contains about 9K webpage screenshots, annotated with regions and types of layout elements, together with the information about the webpage layout type (credential required or not) and the corresponding brand.

\subsubsection*{Layout Elements Extraction}

The \detr (Detection Transformer)~\cite{detr_2020} architecture revolutionizes object detection by employing transformers, to directly learn the relationships between objects in an image without predefined anchor boxes.
Considering its state-of-the-art performance in object detection tasks, we have decided to adopt the \detr model with \tool{ResNet-50}~\cite{resnet50} backbone.
We trained the model using \textit{8,109} labeled webpage screenshots.
During the training stage, we conducted a comparative analysis of the \detr model's performance across varying image sizes.
Table \ref{tab:detr_performance} shows the mAP (Mean Average Precision) accuracy metric compared with processing speed. 
The model which works with small images shows good performance, albeit with lower accuracy.

\begin{table}[h!]
\centering
\caption{\tool{\footnotesize{DETR}} Performance Comparison}
\label{tab:detr_performance}
\begin{tabular}{lccc}
\toprule
\textbf{Image Size} & \textbf{Logo mAP [.5:.95]} & \textbf{Samples per Second} \\
\midrule
224x386 & 37.3 & 10.5 \\
432x768 & 46.6 & 3.9 \\
\bottomrule
\end{tabular}
\end{table}

To safeguard user security, it's imperative that our system processes webpage layout analysis at a minimum rate of once per second.
Concurrently, it's essential to manage the workload on the CPU effectively to prevent overutilization, ensuring the system remains responsive and reliable over extended periods of operation.
Achieving an average processing rate of 3.9 samples per second provides the capability to analyze a frame in approximately 1/4 seconds.
This rapid processing affords us a substantial interval of 3/4 seconds where the CPU can enter a reduced activity state, thereby mitigating the risk of overheating and sustaining performance over time.

\subsubsection*{Reducing Overlapping Boxes}

\begin{figure}[hb!]
    \centering
    \includegraphics[width=\columnwidth]{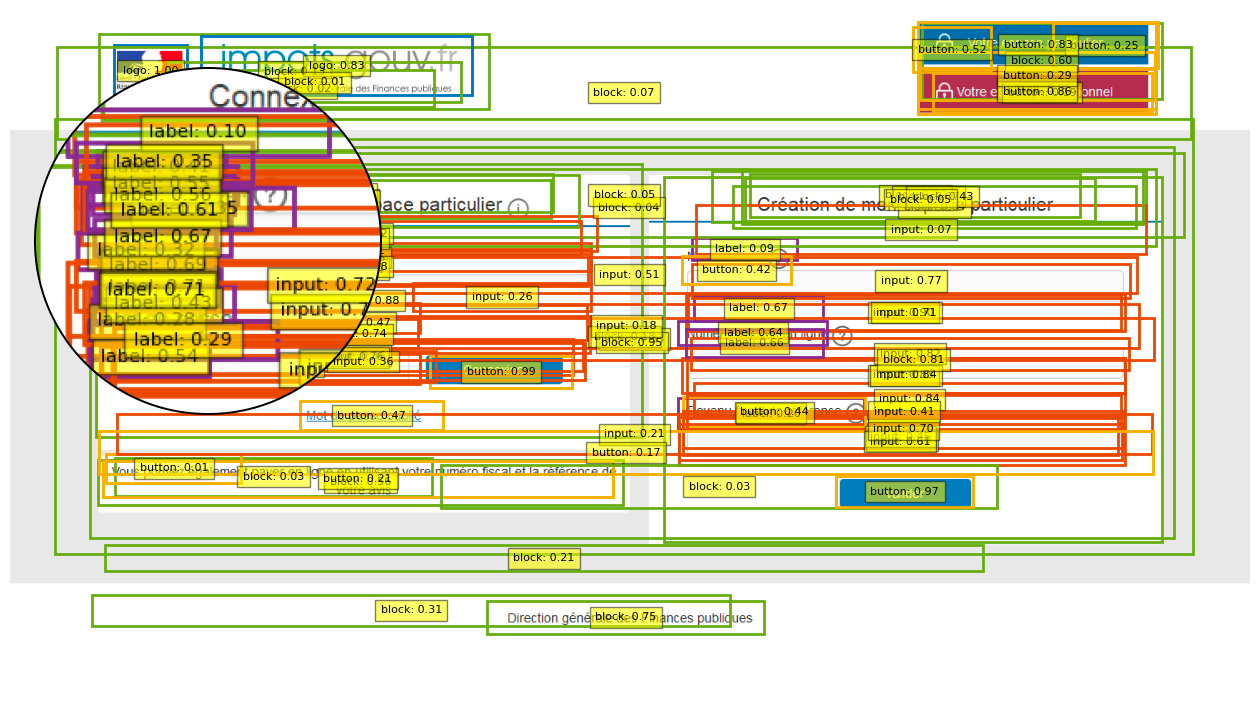}
    \vspace{-10mm}
    \caption{Potential object locations}
    \label{fig:detr-outputs}
\end{figure}

The NMS (Non-Maximum Suppression) algorithm is used in object detection to eliminate redundant or overlapping bounding boxes.
By design, the \detr model doesn't have a built-in NMS layer.
However, given that an abstract webpage layout should never contain overlapping elements, we integrated the \tool{protobuf} model from \tool{coremltools} \footnote{\url{https://apple.github.io/coremltools/docs-guides}}
and merged it with the main model.
\tool{DETR} consistently generates a fixed set of 100 bounding boxes for any input image, which is the default value but can be changed.
These bounding boxes are a representation of potential object locations within the image (Figure~\ref{fig:detr-outputs}).

\begin{figure}[ht!]
    \centering
    \includegraphics[width=\columnwidth]{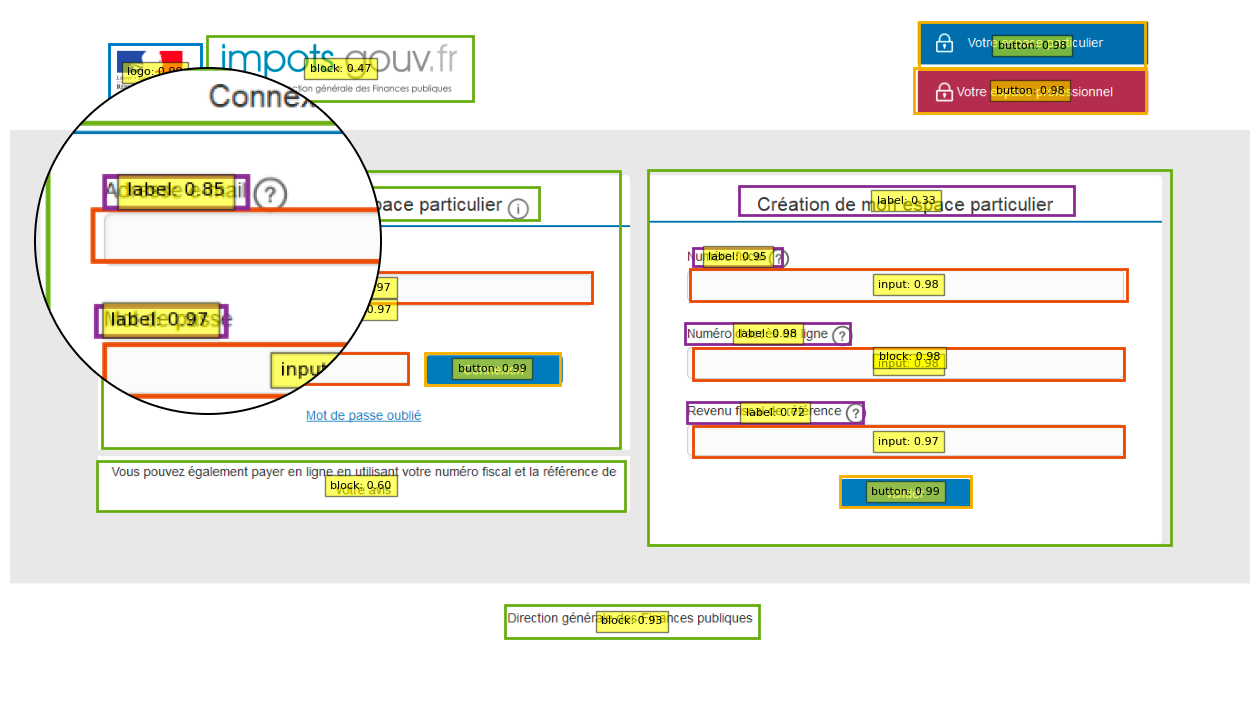}
    \vspace{-10mm}
    \caption{Refined object locations after post-processing}
    \label{fig:detr-nms-outputs}
\end{figure}

NMS systematically identifies and keeps only the non-overlapping bounding boxes with the highest confidence scores and makes our outputs flexible, meaning now we can receive from 0 to 100 boxes instead of a fixed count.
The remaining bounding boxes represent the most likely object locations within the image (Figure~\ref{fig:detr-nms-outputs}).

\subsection{Logo Brand Classification}
\label{ssec:logo_brand_classification}

To train logo brand classification model we used brand logos available in the \textit{OCR-aided Siamese model} dataset part of~\cite{phishintention_dataset}. 
This part contains \textit{3061} logo images of \textit{277} brands.
We utilized the \tool{ResNet-18} model, which was trained on precisely cropped logo layout elements during the training stage.
During the evaluation stage, we utilize the layout elements identified by the \textit{LayoutObjectDetector} (\tool{DETR}) and select the logo element with the highest confidence score.
Subsequently, the \textit{LogoBrandClassifier} (\tool{ResNet-18}) is applied within the bounds of the detected logo's bounding box.
Accuracy metrics details are available in the Section~\ref{sssec:logo_brand_classification_evaluation}.

\subsection{Credential Required Page Classification}
\label{ssec:crp_classification}

\begin{figure}[ht!]
    \centering
    \includegraphics[width=\columnwidth, height=6cm, keepaspectratio]{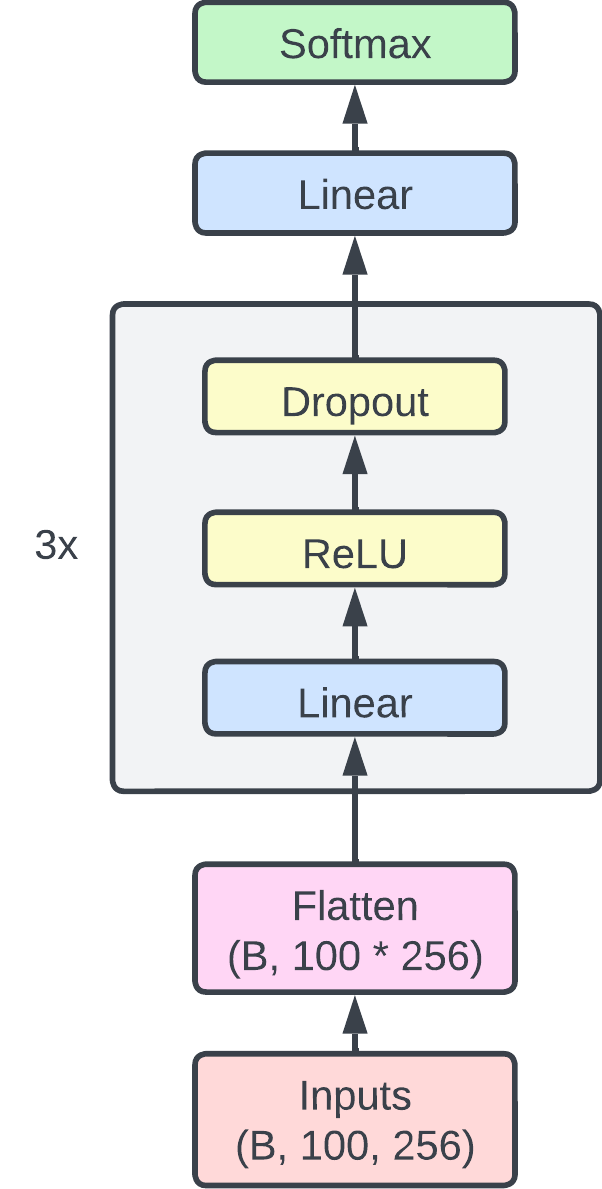}
    \caption{Classification head for {\scriptsize{\textsf{DETR}}} object-level features}
    \label{fig:crp-classifier-layers}
\end{figure}

Credential required page classification is an important step to reduce the number of false positive detections.
To streamline our process and optimize the algorithm we aimed to utilize already extracted features from the Abstract Layout Detection step.
The \detr model receives the query image as an input and generates \textit{N} output vectors.
Because output vectors are used to predict the class and location of the object, they consist of the information about the object and are suitable for the object-level image representation~\cite{ban_image_2022}.
The object detection model was divided into two models, the first is \textit{LayoutFeatureExtractor}, which generates object-level features of shape \textit{(B, N, H)} where \textit{B} is batch size, \textit{N} is a number of output vectors which is 100 in our case and \textit{H} is hidden states for each output which is 256 in our case.
The subsequent \textit{LayoutObjectDetector} model takes these raw features as input and transforms it into a tensor of shape \textit{(B, N, 5)} where the first 4 elements are coordinates of the bounding boxes and the last element is a layout element class.
Then, we employ an additional \textit{LayoutClassifier} model, which repurposes the previously extracted object-level features to transform the task of object detection into image classification.
This transition is facilitated by a custom classification head (Figure~\ref{fig:crp-classifier-layers}), comprising a streamlined multi-layer perceptron architecture.
The classification head processes and flattens the object-level features, applies ReLU activation functions for non-linear processing, and concludes with a Dropout layer to mitigate overfitting. 
Resulting in the output of shape \textit{(B, 2)}, this classification head efficiently provides classification scores, distinctly categorizing pages as credential required or not.
Accuracy metrics details about this approach are available in the Section~\ref{sssec:crp_classification_evaluation}.

\subsection{Models Hierarchy}

\begin{figure}[ht!]
    \centering
    \includegraphics[width=0.9\columnwidth, keepaspectratio]{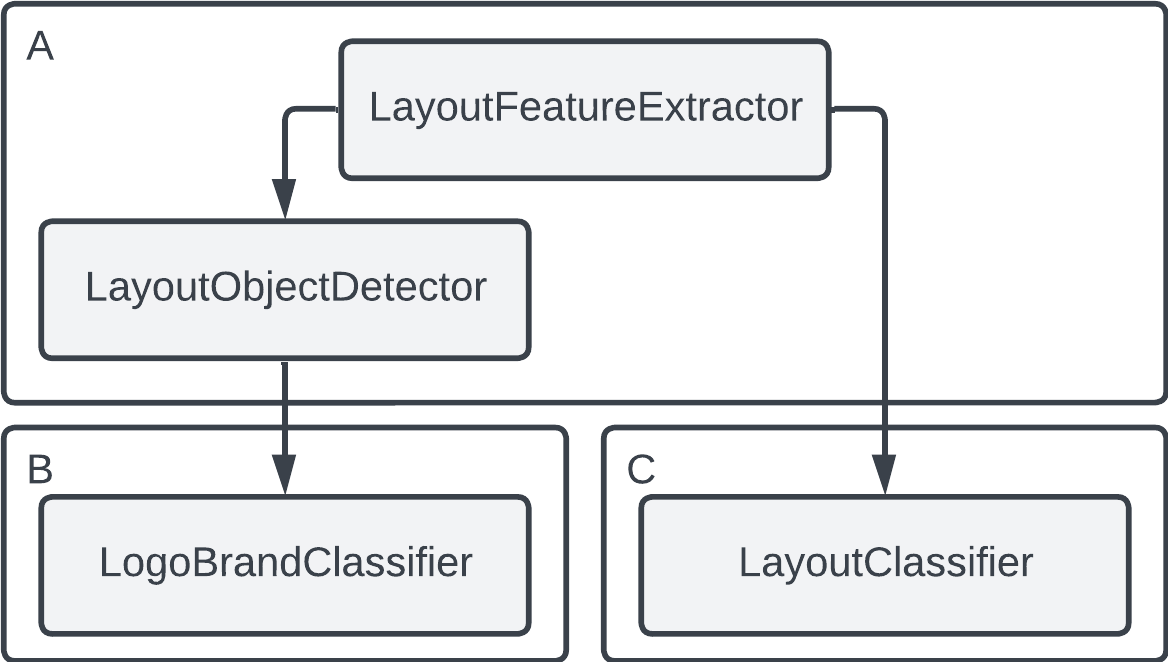}
    \caption{Final Models Hierarchy}
    \label{fig:final-models-hierarchy}
\end{figure}

The resulting hierarchy of the final models is illustrated in Figure~\ref{fig:final-models-hierarchy}, showing the structured organization and interrelations among the different models. 
\textit{A, B} and \textit{C} sections represents different training stages.
\textit{A} is a initial stage where we are training \textit{LayoutFeatureExtractor} simultaneously with \textit{LayoutObjectDetector} on object detection task.
\textit{B} and \textit{C} can be trained in parallel, but only when the previous stage was fully trained.
When we make changes in the \textit{A} stage, then stage \textit{C}, which includes \textit{LayoutClassifier}, should be retrained because it fully depends on \textit{LayoutFeatureExtractor} outputs.

All models were converted to the \coreml format to optimize performance through the use of \tool{Metal Performance Shaders}.
\coreml
\footnote{\url{https://docs.developer.apple.com/documentation/coreml}} %
is a built-in high-level \macos native framework to integrate machine learning models into Apple platform applications.
It is optimized to perform on-device operations by leveraging the CPU, GPU, and Neural Engine chip while minimizing its memory footprint and power consumption.
\metal
\footnote{\url{https://docs.developer.apple.com/documentation/metal/}} framework allows applications to directly interact with a device's GPU.
Machine learning applications leverage \metal's computational acceleration for both training and inference tasks.

\subsection{macOS Application}

We built a native macOS application using Swift programming language and integrated all converted \coreml packages into it.
We utilized \screncapturekit to access fullscreen frames.
\screncapturekit
\footnote{\url{https://docs.developer.apple.com/documentation/screencapturekit}}
is a framework for screen capturing on \macos.
This framework has been designed with a focus on performance by GPU utilization.
It allows capturing content from various sources such as displays, applications, and windows, along with associated audio. 
A native macOS application continuously runs in the background, waiting for the active web browser to appear on the screen.
While working with different browsers, we need to specify the region of actual web content and ignore the desktop background, address input, or tabs.
The \tool{Accessibility Framework}
\footnote{\url{https://docs.developer.apple.com/documentation/accessibility/}}
incorporates an extensive array of tools and functionalities aimed at enhancing the accessibility of Apple devices.
For developers, this framework offers the opportunity to delve into the accessibility graph that contains all UI elements and their relations, providing an additional source of information regarding the active application's state, layout, and controls.
In our case, we can inspect the accessibility elements graph of Safari and find an element called \textit{AXWebArea}, which contains the actual webpage. 
By getting its parent, which is \textit{AXScrollArea}, we can account for the relative coordinates of the web area to the inner page scroll offset.
These coordinates are used to crop the fullscreen image and resize it to 432x768 size, making the image acceptable for the machine learning models analysis.
After this, the algorithm works exactly as described in Section~\ref{sec:solution}.

\section{Evaluation}

We perform a case study, exploring several use aspects to ensure that our system is optimized for both performance and efficiency:

\noindent
\textbf{Abstract Layout Detection Accuracy.} How does our model's logo-detecting efficiency compare to the previous solutions in terms of both accuracy and processing speed?

\noindent
\textbf{Logo Brand Classification Accuracy.} How accurately can we classify the brand of detected logo element?

\noindent
\textbf{Credential Required Page Classification Accuracy.} What is the effectiveness of our object-level features-based classification compared to previous work?

\noindent
\textbf{Overall Phishing Detection Performance.} Did we manage to maintain the overall accuracy of phishing webpage detection compared to other solutions?

\noindent
\textbf{System Resources Usage Efficiency}. Can we perform continuous processing in the background without a major impact on user workflow?

\subsection{Accuracy Metrics}

During the accuracy evaluation, we used the same dataset~\cite{phishintention_dataset} provided by PhishIntention team to compare our solution side by side with competitors.
We provide detailed information about each used dataset part at the beginning of corresponding subsection.

\subsubsection{Abstract Layout Detection}

We used 901 labeled webpages from \textit{AWL detector} dataset part of~\cite{phishintention_dataset} to measure how accurate we can find logo elements.
The mAP (Mean Average Precision) measurement is under the IoU (Intersection over Union) thresholds [0.5:0.95] and shows slightly worse accuracy compared to the PhishIntention model, but at the same time, achieves the same results as Phishpedia~\cite{lin_phishpedia_2021} model (Table~\ref{tab:detr_logo_accuracy}).
We achieved worse results than competitors, but our solution is more efficient in terms of processing speed.

\begin{table}[h!]
\centering
\caption{Logo detection accuracy comparison}
\label{tab:detr_logo_accuracy}
\begin{tabular}{lcc}
\toprule
\textbf{Solution} & \textbf{Logo mAP [.5:.95]} & \textbf{Samples per Second} \\
\midrule
PhishIntention & 59.5 & 0.7 \\
Phishpedia & 46.6 & - \\
Ours & 46.6 & 3.9 \\
\bottomrule
\end{tabular}
\end{table}

\subsubsection{Logo Brand Classification} 
\label{sssec:logo_brand_classification_evaluation}

We used 2000 logo images from \textit{OCR-aided Siamese model} dataset part of ~\cite{phishintention_dataset} to measure how accurate we can classify logo brands.
Table~\ref{tab:logo_classification} presents the accuracy metrics achieved by employing the \tool{ResNet-18} model for logo brand classification. 
We achieved the same level of accuracy as PhishIntention and slightly improved accuracy compared to Phishpedia.

\begin{table}[h!]
\centering
\caption{Logo Brand Classification Accuracy}
\label{tab:logo_classification}
\setlength\tabcolsep{12pt}
\begin{tabular}{lc}
\toprule
\textbf{Solution} & \textbf{Test Accuracy} \\
\midrule
PhishIntention & 89.1 \\
Phishpedia & 83.5 \\
Ours & 90.8 \\
\bottomrule
\end{tabular}
\end{table}

\subsubsection{Credential Required Page Classification} 
\label{sssec:crp_classification_evaluation}

We used 901 labeled webpages from \textit{CRP classifier} dataset part of~\cite{phishintention_dataset} to validate the accuracy and robustness of object-level features based classification (Table \ref{tab:crp_accuarcy}).
The increment (+3.1\%) in test accuracy is particularly noteworthy as it indicates a reduction in false positive detections, which is clearly observed in Figure \ref{fig:fpr-comparison}.

\begin{table}[htbp]
\centering
\caption{CRP Classification Accuracy Comparison}
\label{tab:crp_accuarcy}
\begin{tabular}{lcc}
\toprule
\textbf{Solution} & \textbf{Train Accuracy} & \textbf{Test Accuracy} \\
\midrule
PhishIntention & 99.3 & 95.0 \\
Ours & 99.5 & 98.1 \\
\bottomrule
\end{tabular}
\end{table}

\subsubsection{Overall Phishing Detection} \label{sssec:overall_phishing_detection}

\begin{figure}[ht!]
    \centering
    \includegraphics[width=0.9\columnwidth, keepaspectratio]{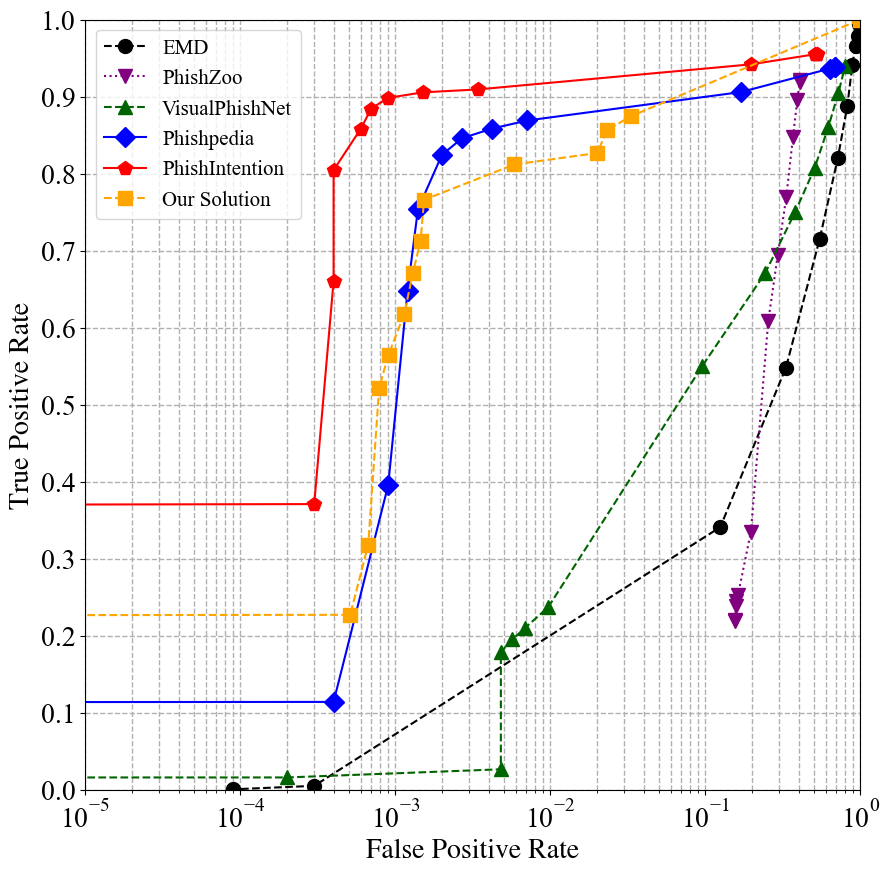}
    \caption{ROC of different phishing detection solutions}
    \label{fig:roc-comparison}
\end{figure}

Figure \ref{fig:roc-comparison} shows the ROC (Receiver operating characteristic) of different phishing detection solutions in logarithm scale.
We evaluated the algorithm within \textit{Experiment dataset (25K benign and 25K phishing webpages)} part of~\cite{phishintention_dataset}.
This plot shows that our solution maintained phishing detection efficiency while bringing it directly to the device.

\begin{figure}[ht!]
    \centering
    \includegraphics[width=0.9\columnwidth, keepaspectratio]{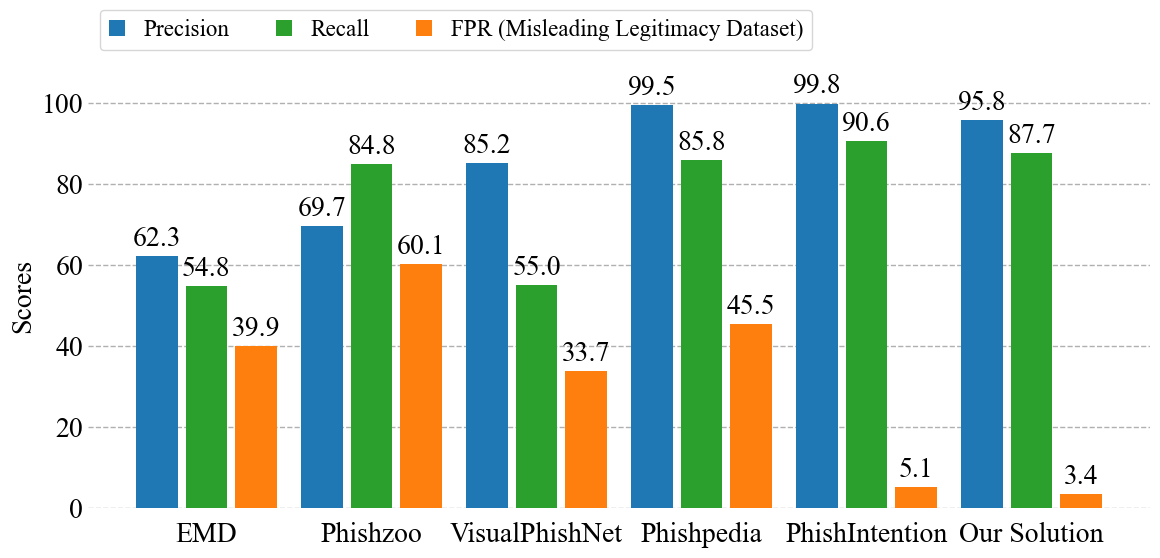}
    \caption{Accuracy metrics of different phishing detection solutions}
    \label{fig:fpr-comparison}
\end{figure}

Figure \ref{fig:fpr-comparison} shows precision and recall within the same dataset, and additionally false positive rate within \textit{Misleading legitimacy} dataset part of~\cite{phishintention_dataset} which contains 3049 benign webpages.
In this plot we can observe that increased accuracy of credential required page classification reduced the false positive rate from 5.1\% to 3.4\%.

\subsection{System Resources Usage}

We measured performance on two different devices: MacBook Pro 2020 M1 and 4K iMac 2020 Intel Core i7 3.8 GHz.
Initial tests were performed without imposing a frame rate limit, enabling the application to analyze each frame as it becomes available.
The results shown in Table~\ref{tab:system_performance} highlight the performance difference between Apple's M1 and Intel's Core i7 processors, with the former demonstrating better optimization in terms of both CPU usage and RAM consumption.

\begin{table}[htbp]
\centering
\caption{System Resources Usage}
\label{tab:system_performance}
\begin{tabular}{lcc}
\toprule
\textbf{Device} & \textbf{CPU Usage} & \textbf{RAM Consumption} \\
\midrule
Apple M1 & 58\% & 440 MB \\
Intel Core i7 & 74\% & 860 MB \\
\bottomrule
\end{tabular}
\end{table}

To meet the criteria for unobtrusive background processing, we restrict the frame rate to a single frame per second.
This measure is posited to strike an optimal balance between enhancing user security and minimizing the consumption of system resources. 
We observe (Table~\ref{tab:system_performance_lock}) a significant improvement in CPU utilization attributed to increased periods of processor idleness.

\begin{table}[htbp]
\centering
\caption{System Resources Usage (FPS Lock)}
\label{tab:system_performance_lock}
\begin{tabular}{lcc}
\toprule
\textbf{Device} & \textbf{CPU Usage} & \textbf{RAM Consumption} \\
\midrule
Apple M1 & 20\% & 440 MB \\
Intel Core i7 & 11\% & 820 MB \\
\bottomrule
\end{tabular}
\end{table}

Quantization\footnote{\url{https://huggingface.co/docs/optimum/concept_guides/quantization}} is a process of clamping the model weights.
We quantized all models from a default 32-bit floating point to a smaller but less accurate 16-bit floating point. 
Model quantization yields substantial enhancements in frame processing time, CPU utilization, bundle size, and, most notably, RAM consumption (Table~\ref{tab:system_performance_quantized}).

\begin{table}[htbp]
\centering
\caption{System Resources Usage (Quantized Models)}
\label{tab:system_performance_quantized}
\begin{tabular}{lcc}
\toprule
\textbf{Device} & \textbf{CPU Usage} & \textbf{RAM Consumption} \\
\midrule
Apple M1 & 16\% & 84M \\
Intel Core i7 & 8\% & 450M \\
\bottomrule
\end{tabular}
\end{table}

Subsequently, we checked how quantization affects overall phishing detection efficiency in the same way as described in Section \ref{sssec:overall_phishing_detection}.
We found that this optimization technique effectively halves the model's size while maintaining a minimal impact on its accuracy (Table~\ref{tab:phishing_detection_efficiency_comparison}).

\begin{table}[htbp]
\centering
\caption{Quantization Impact on Accurracy}
\label{tab:phishing_detection_efficiency_comparison}
\begin{tabular}{lcc}
\toprule
\textbf{Floating-point} & \textbf{Precision} & \textbf{Recall} \\
\midrule
32-bit & 95.8\% & 87.8\% \\
16-bit & 95.1\% & 87.1\% \\
\bottomrule
\end{tabular}
\end{table}

\section{Related Work}
\label{sec:relwork}

\subsubsection*{Reference-based phishing webpage detection} %
State-of-the-art approaches, such as PhishIntention~\cite{liu_inferring_2022}, Phishpedia~\cite{lin_phishpedia_2021}, and VisualPhishNet~\cite{abdelnabi_visualphishnet_2020},
primarily focus on automatically identifying phishing websites on the internet using web crawlers rather than relying on local inference.
Solutions, such as DynaPhish~\cite{2023DynaPhish} and KnowPhish Detector~\cite{2024KnowPhish}, further improve accuracy and extend the applicability of reference-based approaches but still are cloud-first.
Our work addresses the performance aspect of such approaches, bringing them locally to the user devices without compromising on phishing detection accuracy.

\subsubsection*{On-device phishing detection} %
Both industry practitioners and academia researchers have recently begun to explore this attack vector more actively, though, to the best of our knowledge, the current industry focus is on SMS phishing (\emph{smishing}) prevention.
Recent works from Samsung R\&D Institute focus on smishing: Harichandana~\etal~\cite{2024smishingCOPS} introduce an on-device pipeline for smishing detection,
and J.~W.~Seo~\etal~\cite{2024smishingSamsung} present a privacy-preserving SMS classifier.
The privacy-preserving aspect of on-device solutions is also explored by researchers in recent works on SMS spam and phishing detection~\cite{sriraman_-device_2022,2024NextGenSmishing}.

\section{Discussion}
\label{sec:discussion}

\subsubsection*{Proactive Protection}
Our solution aims to automatically detect phishing attacks, eliminating the need for additional user interaction.
There's no need for users to press extra buttons or copy and paste suspicious URLs for detection.
In stressful situations, users may lack the necessary attention to analyze URLs and page content thoroughly. 
Hence, our solution operates proactively, safeguarding users without their active intervention.

\subsubsection*{On-device Processing}
Our final metrics of system resource usage indicate that our solution can operate continuously in the background without a noticeable impact.
It is important to highlight that CPU usage represents a fraction of the total available capacity:
given that the Apple M1 has 8 cores, this means we are utilizing 16\% of the 800\% total capacity available across all cores.
In comparison to other applications, this level of resource consumption is equivalent to maintaining three active Safari browser tabs (which typically consume about 5\% depending on the website) or conducting an active Zoom call (which can use about 22\% during a voice call and about 41\% during video call).

\subsubsection*{Browser Extensions}
In our approach, we have access to the image of the whole screen but select a \emph{region of interest}, which we currently limit to an active browser, performing any screen recognition tasks within this window.
In the use case of browser window content tracking, this approach has a direct advantage over relying on browser extensions, as it is browser-agnostic and requires minimal adjustment for working with different browsers: adding Google Chrome support on the Safari-compatible base required up to 50 lines of code only.
For the users, the browser-agnostic approach has the direct advantage of lessening the burden of browser extension management.

\subsubsection*{Users as Sensors}
Our solution combines with the traditional method of keeping a blacklist of phishing websites.
The algorithm can be extended to perform URL matching with a blacklist at the very beginning.
If there is no match in the blacklist, we can perform the main part of the algorithm described in this paper.
Upon identifying a phishing attempt, both the user and our centralized system are alerted.
The primary advantage of this method is that it leverages thousands of users as sensors rather than relying on multiple web crawlers to search for phishing attempts on the web, %
aligning with the human-as-a-security-sensor (HaaSS)~\cite{2018HaaSS} approach.

\section{Limitations}

The main limitations of our proposed solution stem from the specifics of the use of computer vision techniques for object detection on the webpage.
Our approach only identifies phishing websites that display both a visible brand logo and a credential form on the same page.
If either of these elements is missing or becomes hidden when a user scrolls or magnifies, our solution will fail.

Our solution doesn't protect users from phishing attacks for unknown brands.
In case of the need for a well-known brand logos database update, machine learning models should be retrained and updated on the user device.
Such an approach also fails when the company rebrands.
For example, from Facebook to Meta or from Twitter to X.
A recent study~\cite{lee_attacking_2024} has also shown a possible attack on techniques that rely on brand logo recognition, for which we are not robust.

\section{Conclusions and Future Work}

This work argues for the potential of on-device, real-time phishing detection systems to enhance cybersecurity defensive technologies.
We demonstrate that modern devices are capable of leveraging machine learning models efficiently, sustaining fast processing speed while conserving system resources.
A straightforward future work direction is adapting our solution for iOS devices.
This transition is expected to be relatively smooth, as the \coreml packages required for the transfer are fully compatible with the iOS platform.

In our case study, we focused on webpage phishing detection.
However, given our capability to capture the entire screen, our potential for advancement is not limited here.
A complex solution could orchestrate various algorithms tailored to specific \emph{Regions of Interest.}
For example, in web browsers, we can process webpages according to the introduced algorithm, in email clients, we can detect phishing emails, in messengers, we can detect smishing.

\section{Acknowledgements}

We would like to thank the anonymous reviewers for providing valuable feedback and suggestions that helped to improve this paper.

\bibliographystyle{plain}
\bibliography{main.bbl}

\appendix

\subsubsection*{Object-level Features Based Classification}
We tried to make \textit{BrandClassifier} similar to the \textit{LayoutClassifier} and reuse DETR object-level features to transform it into a tensor of shape (B, 16), which should have probabilities of belonging logo to some brand (Other or Apple, Amazon, Netflix, etc.). 
This approach, while yielding a high precision, suffers from a significantly lower recall rate.
To address this issue, we experimented with the Oversampling and Undersampling techniques.
Oversampling involves increasing the number of instances in the class with fewer instances, while Undersampling reduces the number of instances in the class with more instances.
We observed a significant improvement in the model's overall performance (Table \ref{tab:logo_classification_techniques}), as evidenced by an F1 score that reached 90\%.
However, due to limited accuracy, we decided not to use this model in the final solution.
\begin{table}[h!]
\centering
\caption{Logo Brand Classification Accuracy across different training techniques}
\label{tab:logo_classification_techniques}
\setlength\tabcolsep{12pt}
\begin{tabular}{lccc}
\toprule
\textbf{Technique} & \textbf{F1} & \textbf{Precision} & \textbf{Recall} \\
\midrule
Normal & 0.83 & 0.95 & 0.79 \\
Oversampling & 0.79 & 0.77 & 0.87 \\
Undersampling & 0.90 & 0.98 & 0.86 \\
\bottomrule
\end{tabular}
\end{table}

\subsubsection*{Webpage Screenshot Analysis Example}

\begin{figure*}[hb!]
    \centering
    \includegraphics[width=0.96\textwidth, keepaspectratio]{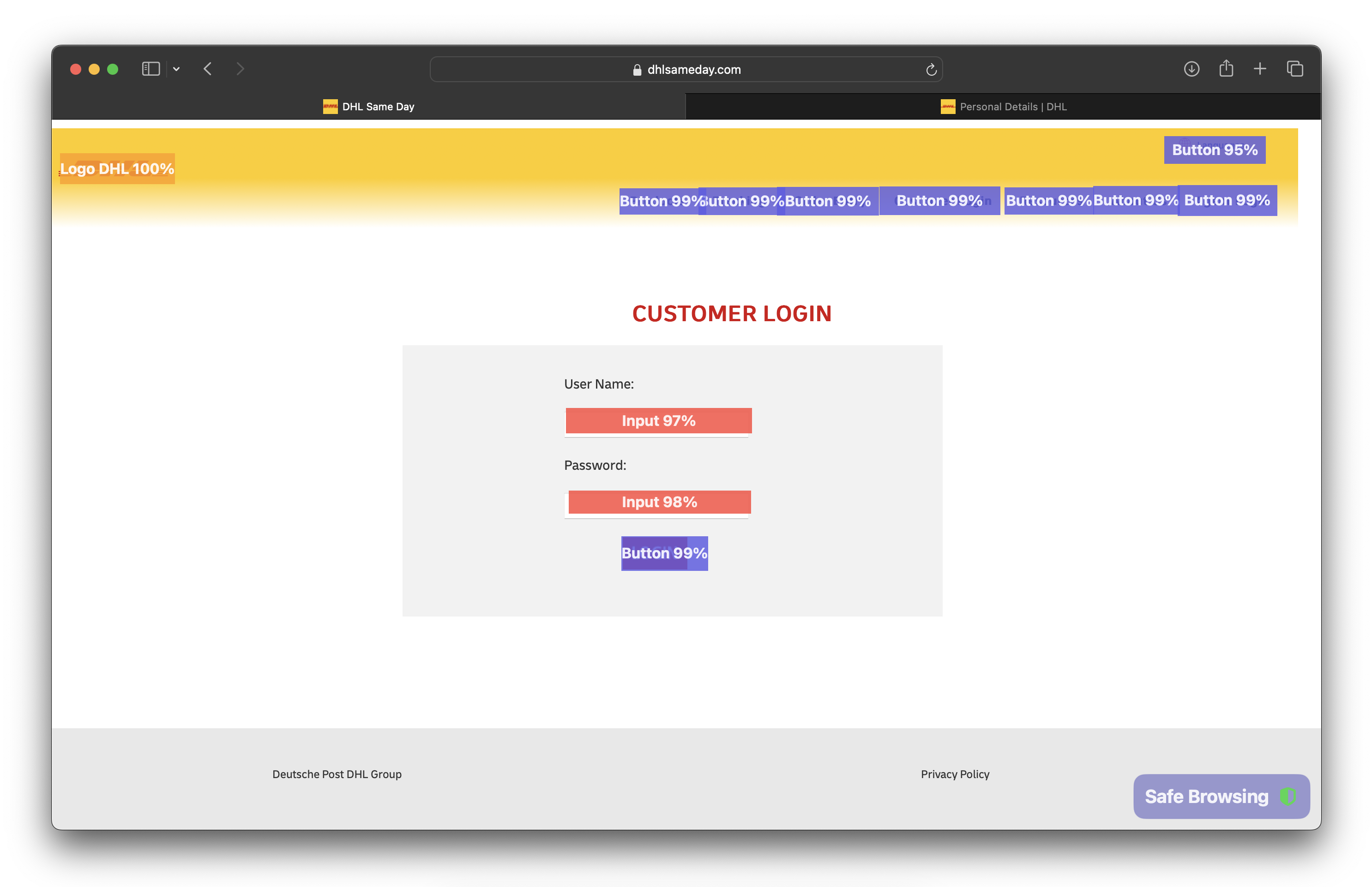}
    \caption{Official DHL website screenshot. This example shows the analysis of the official DHL website, showing the identification of various elements. 
Every button and input was accurately detected, and the logo was confidently identified as belonging to DHL. 
The website layout requires credentials, but the authenticity of dhlsameday.com verifies that it is not a phishing site.}
    \label{fig:dhl-nonphishing-showcase}
\end{figure*}

\begin{figure*}[ht!]
    \centering
    \includegraphics[width=0.96\textwidth, keepaspectratio]{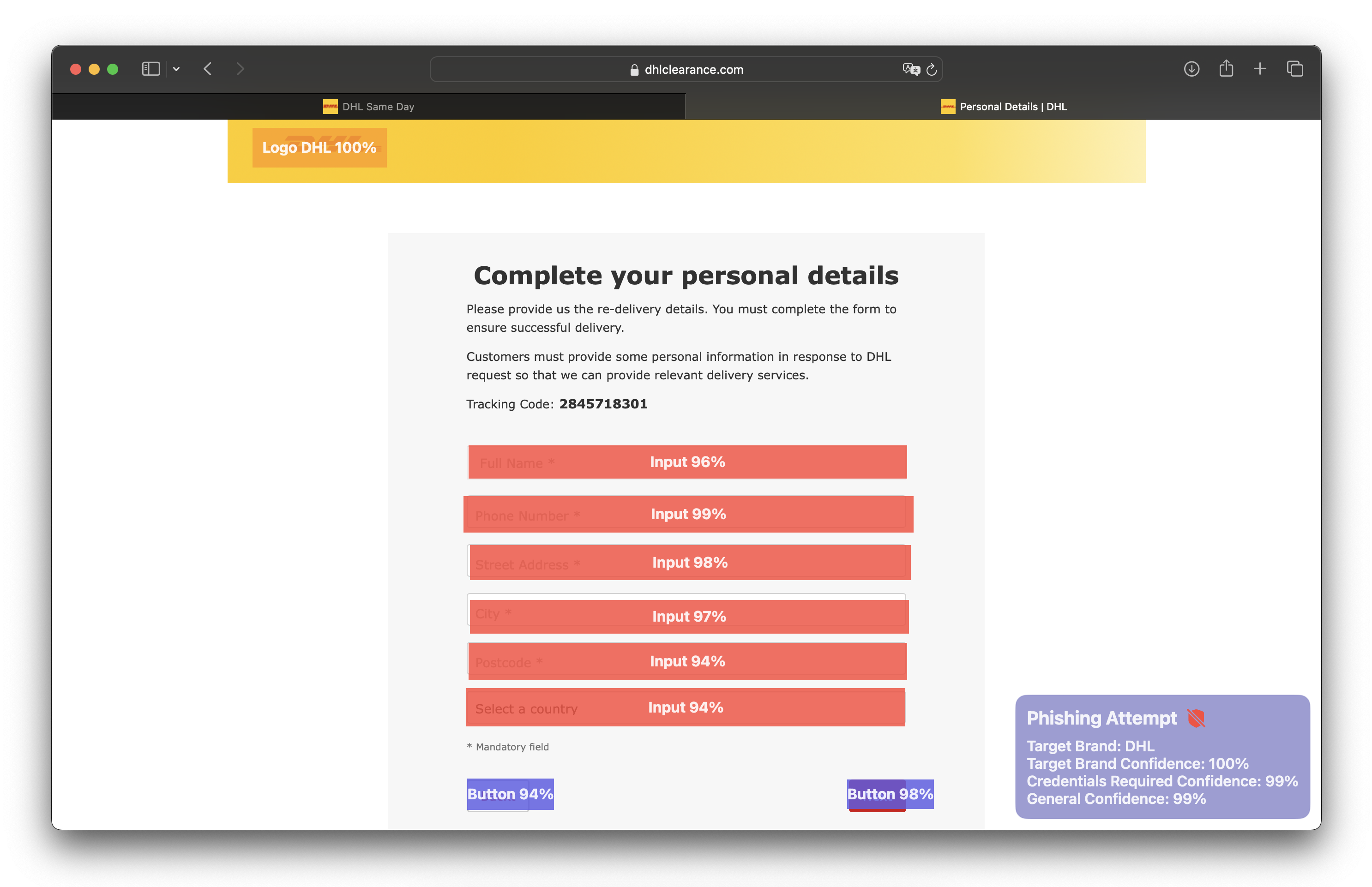}
    \caption{Phishing DHL webpage screenshot. %
    Here, we encountered a layout markedly distinct from the previous example. 
Nevertheless, our evaluation methodically identified all buttons and inputs, and the layout classifier categorized this page as credential required. 
The detection of a logo was a key finding, with the classifier confirming with 100\% confidence its affiliation with DHL. 
However, a closer inspection revealed a critical oversight: the site is not the official DHL webpage but a phishing attempt.
    }
    \label{fig:dhl-phishing-showcase}
\end{figure*}

In Figures~\ref{fig:dhl-nonphishing-showcase} and~\ref{fig:dhl-phishing-showcase}, we illustrate the analysis of the official company webpage and phishing webpage detection by our solution.

\end{document}